\documentclass[journal]{IEEEtran}
\usepackage{graphicx,amssymb,amsmath}
\usepackage{multicol}
\usepackage[noadjust]{cite}
\usepackage{setspace}
\usepackage{stfloats}
\usepackage{amsthm}
\usepackage{flushend}
\usepackage{cases,subeqnarray}
\usepackage{bm,multirow,bigstrut}
\usepackage{textcomp}
\usepackage{latexsym,bm}
\usepackage{booktabs,changebar}
\usepackage{xcolor}
\usepackage{mathtools}
\usepackage{dsfont}
\usepackage{extarrows}
\theoremstyle{plain}

\newtheorem{lemm}{Lemma}
\theoremstyle{plain}
\newtheorem{rem}{Remark}

\IEEEoverridecommandlockouts
\begin{document}
\title{On the Spectral Efficiency of Massive MIMO Systems with Low-Resolution ADCs}
\author{Jiayi~Zhang,~\IEEEmembership{Member,~IEEE,}
        Linglong~Dai,~\IEEEmembership{Senior~Member,~IEEE,}
        Shengyang~Sun,
        and~Zhaocheng~Wang~\IEEEmembership{Senior~Member,~IEEE,}
\vspace{-0.7cm}
\thanks{
This work was supported in part by the International Science \& Technology Cooperation Program of China (Grant No. 2015DFG12760), the National Natural Science Foundation of China (Grant Nos. 61571270 and 61271266),  the Beijing Natural Science Foundation (Grant No. 4142027), and the Foundation of Shenzhen government.}%
\thanks{J. Zhang is with the School of Electronics and Information Engineering, Beijing Jiaotong University, Beijing 100044, P. R. China (e-mails: jiayizhang@bjtu.edu.cn).}
\thanks{L. Dai, S. Sun and Z. Wang are with Department of Electronic Engineering as well as Tsinghua
National Laboratory of Information Science and Technology (TNList), Tsinghua University, Beijing 100084, P. R. China.}
}

\maketitle

\begin{abstract}
The low-resolution analog-to-digital convertor (ADC) is a promising solution to significantly reduce the power consumption of radio frequency circuits in massive multiple-input multiple-output (MIMO) systems. In this letter, we investigate the uplink spectral efficiency (SE) of massive MIMO systems with low-resolution ADCs over Rician fading channels, where both perfect and imperfect channel state information are considered. By modeling the quantization noise of low-resolution ADCs as an additive quantization noise, we derive tractable and exact approximation expressions of the uplink SE of massive MIMO with the typical maximal-ratio combining (MRC) receivers. We also analyze the impact of the ADC resolution, the Rician $K$-factor, and the number of antennas on the uplink SE. Our derived results reveal that the use of low-cost and low-resolution ADCs can still achieve satisfying SE in massive MIMO systems.
\end{abstract}


\IEEEpeerreviewmaketitle
\vspace{-0.2cm}
\section{Introduction}
Recently,  massive multiple-input multiple-output (MIMO) has attracted significant interests for 5th generation (5G) wireless systems, as it is able to achieve dramatic gain in spectral efficiency (SE) \cite{ngo2013energy,gao2015spatially}.
In receiver radio-frequency (RF) circuits, massive analog-to-digital converters (ADCs) corresponding to massive antennas are used to convert the received RF signal to the baseband. A typical flash ADC with $b$-bit resolution and the sampling frequency $f_s$ operates $f_s 2^b$ conversion steps per second, which means the power consumption of ADCs scales exponentially with the resolution and linearly with the sampling rate \cite{walden1999analog}.
Therefore, ADCs with high-speed (e.g., 1 GSample/s) and high-resolution (e.g., 8-12 bits) will put a heavy burden on the power consumption of massive MIMO systems, which is considered as the bottleneck to realize massive MIMO in practice \cite{gao2015energy}.

To solve the power consumption problem, the time-interleaved ADC can be used to reduce the sampling rate, but such solution causes an inevitable error floor of the system performance. Another promising solution is to employ low-resolution ADCs (e.g., 1-3 bits) at RF chains. Great efforts have been made to understand the effects of low-resolution ADCs on the performance of MIMO systems \cite{orhan2015low,mo2015capacity,risi2014massive,liang2015mixed}.
For conventional small-scale MIMO systems, the impact of ADC resolution on the SE was studied by using an additive quantization noise model (AQNM), which takes into account the dependency of ADC power on the bit resolution \cite{orhan2015low}. The SE of flat fading MIMO channels with one-bit ADCs has been analyzed in \cite{mo2015capacity}, while the impact of one-bit ADCs on the SE of massive MIMO systems was recently investigated in \cite{risi2014massive,liang2015mixed}.
Unfortunately, the aforementioned works only consider Rayleigh fading channels and one-bit resolution. The assumption of Rayleigh fading channels is often violated in an line-of-sight (LoS) path dominated practical wireless propagation scenarios, where the Rician fading model is more accurate, and massive MIMO has found its very promising application in such scenarios \cite{zhang2014power,zhang2015achievable}. {However, the effect of low-resolution ADCs has been ignored in \cite{zhang2014power,zhang2015achievable}.} Moreover, the analysis of the effect of more than one-bit resolution (e.g., 2-3 bits) on massive MIMO systems is still limited. Only recently, \cite{Fan2015uplink} investigated the uplink performance of massive MIMO systems with 1-2 bits resolution ADCs and perfect channel state information (CSI). {While a general analytic framework which accounts for more realistic fading models and imperfect CSI seems to be missing in \cite{Fan2015uplink}.}



In this letter, we focus on the uplink SE of massive MIMO systems with low-resolution ADCs over Rician fading channels. {Our derived expressions present a generalization of previous results, which assume ideal ADCs \cite{zhang2014power} or Rayleigh fading channels \cite{Fan2015uplink}.} With the help of AQNM, we investigate the uplink SE both with perfect and imperfect CSI. The derived exact limits and approximations reveal the effects of ADC resolution, the Rician $K$-factor, and the number of BS antennas on the uplink SE. In addition, the power scaling law of massive MIMO systems with low-resolution ADCs is also presented. More importantly, we have an interesting finding that there appears a fixed constant loss of SE related to the ADC quantization bit, which is quite different from unquantized massive MIMO systems. Finally, it is shown that the Rician $K$-factor has a noticeable reduction effect on the SE of massive MIMO systems with low-resolution ADCs.


\emph{Notation:} We use upper and lower case boldface to denote matrices and vectors, respectively. The expectation is given by $\mathbb{E}\{\cdot\}$, while the $N \times N$ identity matrix reads as ${\bf{I}}_N$. Let ${\text{tr}}(\bf{X})$, ${\bf{X}}^T$, and ${\bf{X}}^H$ be the trace, transpose, and conjugate transpose of a matrix  $\bf{X}$, respectively. The matrix determinant and trace are given by $|\bf{X}|$ and ${\tt{tr}}(\bf{X})$. Moreover, $\text{diag}(\bf{X})$ keeps only the diagonal entries of $\bf{X}$. Finally, $x_{ij}$ and ${\bf{x}}_{i}$ denotes the $(i, j)$th entry and the $i$th column of a matrix $\bf{X}$, respectively.

\section{System Model}\label{se:model}
We consider a typical massive MIMO system with low-resolution ADCs, where $N$ single-antenna users simultaneously communicate with an $M$-antenna BS using the same time-frequency resources. The uplink received vector ${\bf{y}} \in \mathbb{C}^{M\times 1}$ can be written as
\begin{align}\label{eq:received_signal}
{\bf{y}} = \sqrt{p_u}{\bf{G}}{\bf{x}}+{\bf{n}},
\end{align}
where $p_u$ is the transmit power of each user, ${\bf{x}}\in \mathbb{C}^{N\times 1}$ denotes the transmit vector for all $N$ users, and the elements of ${\bf{n}}$ represent the additive white Gaussian noise (AWGN) with zero-mean and unit variance. We assume the input covariance matrix ${{\bf{R}}_{\bf{x}}} = {{\bf{I}}_N}$. The channel matrix between the BS and users is given by ${\bf{G}} =  {\bf{H}}{\bf{D}}^{1/2}$, where ${\bf{D}}  \in \mathbb{C}^{N \times N}$ is the large-scale fading channel matrix to model both the geometric attenuation and shadow fading with the elements $\beta_n$,
and ${\bf{H}} \in \mathbb{C}^{M \times N}$ denotes the small-scale fading channel matrix including a deterministic component ${\bf{\bar{H}}}$ corresponding to the LoS signal and a Rayleigh-distributed random component ${\bf{{H}}}_\omega$ accounting for the scattered signals as
\begin{align}\label{eq:Rician_matrix}
{\bf{H}} = {\bf{\bar H}}\sqrt {{\bf{\Omega }}{{\left( {{\bf{\Omega }} + {{\bf{I}}_N}} \right)}^{ - 1}}}  + {{\bf{H}}_\omega }\sqrt {{\bf{\Omega }}{{\left( {{\bf{\Omega }} + {{\bf{I}}_N}} \right)}^{ - 1}}},
\end{align}
where ${\bf{H}}_\omega \sim \mathcal{C}\mathcal{N}({\bf{0}},{\bf{I}}_N)$, ${\bf{\bar{H}}}$ have an arbitrary rank as
\begin{align}\label{eq:deterministic_component}
{\left[ {{\bf{\bar H}}} \right]_{mn}} = \exp \left( { - j\left( {m - 1} \right)k\sin {\theta _n}} \right),
\end{align}
where $ {\theta _n}$ denotes the arrival angle of the $n$th user, $k=2\pi d/\lambda$ with $\lambda$ being the wavelength, and $d$ represents the inter-element space.
Furthermore, the elements of the diagonal matrix ${\bf{\Omega }} $ is $K_n$, which denotes the Rician $K$-factor.

As the quantization error can be well approximated as a linear gain with AQNM \cite{orhan2015low,bai2015energy}, the output vector of low-resolution ADCs is given by \cite[Eq. (1)]{orhan2015low}
\begin{align}\label{eq:quantization_error}
{{\bf{y}}_q} = \mathbb{Q}\left\{ {\bf{y}} \right\} = \kappa {\bf{y}} + {{\bf{n}}_q} = \kappa \sqrt {{p_u}} {\bf{Gx}} + \kappa {\bf{n}} + {{\bf{n}}_q},
\end{align}
where $\mathbb{Q}\{\cdot\}$ is the quantizer function, and $\kappa = 1- \rho$ is a linear gain with \cite[Eq. (13)]{fletcher2007robust}
\begin{align}\label{eq:inverse_coding_gain}
\rho  =  {{\mathbb{E}\left\{ {{{\left| {y - {y_q}} \right|}^2}} \right\}}}/ {{\mathbb{E}\left\{ {{{\left| y \right|}^2}} \right\}}}
\end{align}
denoting a proportionality constant between the quantizer input variance and quantizer error variance. As a measure of the ADC's relative accuracy, the exact values of $\rho$ for low bit resolution $b$ can be found in \cite{max1960quantizing}.
With the help of \eqref{eq:quantization_error} and \eqref{eq:inverse_coding_gain}, the covariance matrix of ${{\bf{n}}_q}$ can be written as
\begin{align}\label{eq:error_covariance}
{{\bf{R}}_{{{\bf{n}}_q}}} = \kappa \rho {\text{diag}}\left( {p_u}{{\bf{G}}{{\bf{G}}^H} + {{\bf{I}}_M}} \right),
\end{align}
where $\bf{G}$ is a fixed channel realization.

\newcounter{mytempeqncnt3}
\begin{figure*}[!b]
\normalsize
\setcounter{mytempeqncnt3}{\value{equation}}
\hrulefill
\vspace*{-4pt}
\setcounter{equation}{10}
\begin{align}\label{eq:achievable_SE_MRC_appro_result_P}
 R_{{\text{P}},n}^{{\text{MRC}}} \approx  {\log _2}\left( {1  +  {\frac{{\kappa {p_u}{\beta _n}\left( {{M^2}{{\left( {{K_n} + 1} \right)}^2} + 2M{K_n} + M} \right)}}{{\kappa {p_u}\left( {{K_n} + 1} \right)\sum\limits_{i \ne n}^N {{\beta _i}{\Delta _1}}  + M{{\left( {{K_n} + 1} \right)}^2} + \rho {p_u}M\left( {{\beta _n}\left( {K_n^2 + 4{K_n} + 2} \right) + {{\left( {{K_n} + 1} \right)}^2}\sum\limits_{i \ne n}^N {{\beta _i}} } \right)}}}}\right),
\end{align}
\vspace*{-4pt}
\setcounter{equation}{\value{mytempeqncnt3}}
\end{figure*}


\newcounter{mytempeqncnt2}
\begin{figure*}[!b]
\normalsize
\setcounter{mytempeqncnt2}{\value{equation}}
\vspace*{-4pt}
\hrulefill
\vspace*{-4pt}
\setcounter{equation}{17}
\begin{align}\label{eq:achievable_SE_appro_result_mrc}
R_{{\text{IP}},n}^{{\text{MRC}}} \approx {\log _2}\left( {1 + \frac{{\kappa {p_u}{\beta _n}\left( {{M^2}K_n^2 + 2M{K_n}{\eta _n}(1 + M) + ({M^2} + M)\eta _n^2} \right)}}{{\kappa {p_u}\left( {{K_n} + 1} \right)\sum\limits_{i \ne n}^N {{\beta _i}{\Delta _2}}  + M\left( {{K_n} + {\eta _n}} \right)\left( {{K_n} + 1} \right)\left( {{p_u}\sum\limits_{i = 1}^N {\sigma _i^2}  + 1} \right) + \rho {p_u}{\Delta _3}}}} \right).
\end{align}
\vspace*{-4pt}
\setcounter{equation}{\value{mytempeqncnt2}}
\end{figure*}
\vspace*{-4pt}

\vspace{-2mm}
\section{Uplink Spectral Efficiency }\label{se:uplink}
In this section, we derive the uplink SE of massive MIMO systems with low-resolution ADCs. Without loss of generality, the maximal-ratio combining (MRC) receiver is considered, as it performs fairly well for massive MIMO systems \cite{ngo2013energy}.

We first consider the case of perfect CSI at the BS. The MRC receiver matrix ${\bf{A}}$ depends on the channel matrix ${\bf{G}}$. By multiplying ${{\bf{y}}_q}$ with ${\bf{A}}^H$ and using \eqref{eq:quantization_error}, the quantized output vector for the MRC receiver is given by
\begin{align}\label{eq:quantized_output_vector}
{\bf{r}} = {{\bf{A}}^H}{{\bf{y}}_q} = \kappa \sqrt {{p_u}} {{\bf{A}}^H}{\bf{Gx}} + \kappa {{\bf{A}}^H}{\bf{n}} + {{\bf{A}}^H}{{\bf{n}}_q}.
\end{align}
The output signal for the $n$th user can be expressed as
\begin{align}
{r_n} = \kappa \sqrt {{p_u}} {\bf{a}}_n^H{{\bf{g}}_n}{x_n} + \kappa \sqrt {{p_u}} \sum\limits_{i \ne n}^N {{\bf{a}}_n^H{{\bf{g}}_i}{x_i}}  + \kappa {\bf{a}}_n^H{\bf{n}} + {\bf{a}}_n^H{{\bf{n}}_q}.\notag
\end{align}
Given perfect CSI, the SE of the $n$th user ${R_{{\rm{P}},n}}$ is
\begin{align}\label{eq:uplink_achievable_rate}
{R_{{\text{P}},n}} = \mathbb{E}\left\{ {{{\log }_2}\left( {1 + \frac{{\kappa {p_u}{{\left| {{\bf{a}}_n^H{{\bf{g}}_n}} \right|}^2}}}{\Psi_1 }} \right)} \right\},
\end{align}
where the variance of the interference-plus-noise variable is
\begin{align}
\Psi_1 ={\kappa {p_u}\sum\limits_{i \ne n}^N {{{\left| {{\bf{a}}_n^H{{\bf{g}}_i}} \right|}^2}} \!+ \!{{\left\| {{{\bf{a}}_n}} \right\|}^2} \!+ \!\rho {p_u}{\bf{a}}_n^H{\text{diag}}\left( {{\bf{G}}{{\bf{G}}^H}} \right){{\bf{a}}_n}}.\notag
\end{align}

For the more practical and general case of imperfect CSI, we consider a transmission within the coherence interval $T$ and use $\tau$ symbols for pilots. The power of pilot symbols is $p_p \triangleq \tau p_u$, while the minimum mean-squared error (MMSE) estimate of ${\bf{G}}$ is ${\bf{\hat G}}$. Let ${\bf{\Xi }} \buildrel \Delta \over = {\bf{\hat G}} - {\bf{G}}$ be the channel estimation error matrix, which is independent of ${\bf{\hat G}}$. The elements of ${\bf{\Xi }}$ has a variance of $\sigma_n^2 \triangleq{\frac{{{\beta _n}}}{{\left( {1 + {p_p}{\beta _n}} \right)\left( {{K_n} + 1} \right)}}}$ \cite[Eq. (21)]{zhang2014power}. Following a similar analysis in \cite{ngo2013energy}, the uplink SE of the $n$th user under imperfect CSI ${R_{{\rm{IP}},n}}$ is\footnote{{We can utilize the mixed-ADC architecture \cite{liang2015mixed} to perform channel estimation for each antennas in a round-robin manner, and thus the effect of low-resolution ADCs can be omitted in the channel estimation phase.}}
\begin{align}\label{eq:uplink_achievable_rate_IP}
{R_{{\text{IP}},n}} = \mathbb{E}\left\{ {{{\log }_2}\left( {1 + \frac{{\kappa {p_u}{{\left| {{\bf{\hat a}}_n^H{{{\bf{\hat g}}}_n}} \right|}^2}}}{{{\Psi _2}}}} \right)} \right\},
\end{align}
where
\begin{align}\label{eq:Psi_2}
{\Psi _2} &= \kappa {p_u}\sum\limits_{i \ne n}^N {{{\left| {{\bf{\hat a}}_n^H{{{\bf{\hat g}}}_i}} \right|}^2}}  + \kappa {p_u}{\left\| {{{{\bf{\hat a}}}_n}} \right\|^2}\sum\limits_{i = 1}^N {\sigma_i^2}+ \kappa {\left\| {{{{\bf{\hat a}}}_n}} \right\|^2} \notag \\
&+ {\rho {\bf{\hat a}}_n^H{\rm{diag}}\left( {{p_u}\left( {{\bf{\hat G}} - {\bf{\Xi }}} \right){{\left( {{\bf{\hat G}} - {\bf{\Xi }}} \right)}^H} + {\bf{I}}} \right){{{\bf{\hat a}}}_n}}.
\end{align}

With MRC receivers, the linear receiver matrixes for perfect and imperfect CSI are ${\bf{A}}={\bf{G}}$ and ${\bf{\hat A}}={\bf{\hat G}}$, respectively.
\vspace{-3mm}
\subsection{Perfect CSI}
For the case of perfect CSI, we have following Lemma 1.
\begin{lemm}\label{prop:mrc}
For massive MIMO MRC systems over Rician fading channels with low-resolution ADCs and MRC receivers with perfect CSI, the approximated SE of the $n$th user is given by \eqref{eq:achievable_SE_MRC_appro_result_P} on the bottom of next page,
where
\setcounter{equation}{11}
\begin{align}\label{eq:notations}
{\Delta _1} &\buildrel \Delta \over = \frac{  \left( {{K_n}{K_i}\phi _{ni}^2 + M\left( {{K_n} + {K_i}} \right) + M} \right) } {\left( {{K_i} + 1} \right)} ,\\
{\phi _{ni}} &\buildrel \Delta \over = \frac{{\sin \left( {{{M\pi \left( {\sin {\theta _n} - \sin {\theta _i}} \right)} \mathord{\left/
 {\vphantom {{M\pi \left( {\sin {\theta _n} - \sin {\theta _i}} \right)} 2}} \right.
 \kern-\nulldelimiterspace} 2}} \right)}}{{\sin \left( {{{\pi \left( {\sin {\theta _n} - \sin {\theta _i}} \right)} \mathord{\left/
 {\vphantom {{\pi \left( {\sin {\theta _n} - \sin {\theta _i}} \right)} 2}} \right.
 \kern-\nulldelimiterspace} 2}} \right)}}.
\end{align}
\end{lemm}
\begin{IEEEproof}
Using \cite[Lemma 1]{zhang2014power} on \eqref{eq:uplink_achievable_rate}, we can derive
\begin{align}\label{eq:MRC_appro}
R_{{\text{P}},n}^{{\text{MRC}}} \approx {\log _2}\left( {1 \!+\! \frac{{\mathbb{E}\left\{ {\kappa {p_u}{{\left\| {{{\bf{g}}_n}} \right\|}^4}} \right\}}}{{\mathbb{E}\left\{  \Psi_1 \right\}}}} \right).
\end{align}
The expectations in \eqref{eq:MRC_appro} can be evaluated with the help of \cite[Lemma 3]{zhang2014power}.
Based on the definition of \eqref{eq:Rician_matrix}, the last term in the denominator of \eqref{eq:MRC_appro} becomes
\begin{align}
&\mathbb{E}\left\{ {{\bf{g}}_n^H{\rm{diag}}\left( {{\bf{G}}{{\bf{G}}^H}} \right){{\bf{g}}_n}} \right\} \notag \\
 &= \sum\limits_{m = 1}^M {\mathbb{E}\left\{ {{{\left| {{g_{mn}}} \right|}^4}} \right\}}  + \sum\limits_{m = 1}^M {\sum\limits_{i \ne n}^N {\mathbb{E}\left\{ {{{\left| {{g_{mn}}} \right|}^2}} \right\}\mathbb{E}\left\{ {{{\left| {{g_{mi}}} \right|}^2}} \right\}} } \notag\\
 &= \beta _n^2M\frac{{K_n^2 + 4{K_n} + 2}}{{{{\left( {{K_n} + 1} \right)}^2}}} + M{\beta _n}\sum\limits_{i \ne n}^N {{\beta _i}}\label{eq:notations_p}.
\end{align}
The proof can be concluded after substituting \eqref{eq:notations_p} into \eqref{eq:MRC_appro} and some simplification.
\end{IEEEproof}

Note that Lemma \ref{prop:mrc} is a general result and reveals the impact of the ADC resolution, the Rician $K$-factor, along with the number of BS antennas on the SE. From \eqref{eq:achievable_SE_MRC_appro_result_P}, the SE increases with more resolution bits. To further reveal the relationship between the SE and key parameters, the following important analysis is provided.

\begin{rem}
For the case of infinite precision ($b\rightarrow \infty$), \eqref{eq:achievable_SE_MRC_appro_result_P} coincides with \cite[Eq. (40)]{zhang2014power} where the quantization error of low-resolution ADCs $\rho$ tends to zero.
When $K_n=K_i=0$, \eqref{eq:achievable_SE_MRC_appro_result_P} reduces to the Rayleigh fading channels as
\begin{align}\label{eq:achievable_SE_MRC_appro_result_P_Rayleigh}
R_{{\rm{P}},n}^{{\rm{MRC}}} \!\to\!  {\log _2}\left( {1 + \frac{{\kappa {p_u}{\beta _n}\left( {M + 1} \right)}}{{{p_u}\sum\limits_{i \ne n} {{\beta _i}}  + 2\rho {p_u}{\beta _n} + 1}}} \right).
\end{align}
For the power-scaling law and assuming $p_u=E_u/M$ with a fixed $E_u$, the large antenna limit of \eqref{eq:achievable_SE_MRC_appro_result_P} converges to
\begin{align}\label{eq:power_scaling_mrc_perfect}
R_{{\rm{P}},n}^{{\rm{MRC}}}  \to  {\log _2}\left( {1  + \kappa {E_u}{\beta _n}} \right) ,\; \text{as} \; M \!\to\! \infty.
\end{align}
Note that \eqref{eq:power_scaling_mrc_perfect} depends on the Rician $K$-factor. As opposed to the ideal ADC, there is a non-zero constant coefficient $\kappa$ when scaling down the transmit power proportionally to $1/M$. Furthermore, it is clear from \eqref{eq:power_scaling_mrc_perfect} that the SE can be improved by using more quantization bits (larger values of $\kappa$).
\end{rem}

\vspace{-5mm}

\subsection{Imperfect CSI}
Next, we will investigate the uplink SE of massive MIMO systems for the case of imperfect CSI. 

\begin{lemm}\label{prop:mrc_ip}
For massive MIMO with low-resolution ADCs and MRC receivers with imperfect CSI, the approximated uplink SE of the $n$th user is given by \eqref{eq:achievable_SE_appro_result_mrc} at the bottom of this page,
where ${\eta _n} \buildrel \Delta \over = \frac{{{p_p}{\beta _n}}}{{1 + {p_p}{\beta _n}}}$, and
\setcounter{equation}{18}
\begin{align}\label{eq:notations_2}
{\Delta _2} &\buildrel \Delta \over = \frac{{{K_n}{K_i}\phi _{ni}^2 + M{K_i}{\eta _n} + M{K_n}{\eta _i} + M{\eta _n}{\eta _i}}}{{{K_i} + 1}}, \\
{\Delta _3} &\buildrel \Delta \over = M{\beta _n}\left( {K_n^2 + 4{K_n}{\eta _n} + 2\eta _n^2} \right) \notag \\
&+ M\left( {{K_n} + 1} \right)\left( {{K_n} + {\eta _n}} \right)\sum\limits_{i \ne n}^N {\frac{{{\beta _i}}}{{{K_i} + 1}}\left( {{K_i} + {\eta _i}} \right)} .
\end{align}
\end{lemm}
\begin{IEEEproof}
With ${{{{\bf{\hat a}}}_n}} = {{{{\bf{\hat g}}}_n}}$ and using \cite[Lemma 1]{zhang2014power} again, \eqref{eq:uplink_achievable_rate_IP} can be written as
\begin{align}\label{eq:MRC_psi_2}
R_{{\rm{IP}},n}^{{\rm{MRC}}} \approx {\log _2}\left( {1 + \frac{{\mathbb{E}\left\{ {\kappa {p_u}{{\left\| {{{{\bf{\hat g}}}_n}} \right\|}^4}} \right\}}}{{\mathbb{E}\left\{ {{\Psi _2}} \right\}}}} \right).
\end{align}
The expectations in \eqref{eq:MRC_psi_2} can be further derived by utilizing the results in \cite[Lemma 5]{zhang2014power}. Similar to \eqref{eq:notations_p}, the last expectation in the denominator of \eqref{eq:MRC_psi_2} can be calculated as
\begin{align}\label{eq:MRC_psi_3}
&\mathbb{E}\left\{ {{\bf{\hat g}}_n^H{\text{diag}}\left( {{\bf{\hat G}}{{{\bf{\hat G}}}^H}} \right){{{\bf{\hat g}}}_n}} \right\}= \notag \\
 &= \frac{{M\beta _n^2}}{{{{\left( {{K_n} + 1} \right)}^2}}}\left( {K_n^2 + 4{K_n}{\eta _n} + 2\eta _n^2} \right) \notag \\
 &+ \frac{{M{\beta _n}}}{{{K_n} + 1}}\left( {{K_n} + {\eta _n}} \right)\sum\limits_{i \ne n}^N {\frac{{{\beta _i}}}{{{K_i} + 1}}\left( {{K_i} + {\eta _i}} \right)}.
\end{align}
Then we can finish the proof after some simplifications.
\end{IEEEproof}
\begin{rem}
For full resolution ADCs as $\rho =0$, our result \eqref{eq:achievable_SE_appro_result_mrc} can reduce to \cite[Eq. (82)]{zhang2014power}. With $p_u=E_u/{M^\alpha}$ and $M \rightarrow \infty$, the limit of \eqref{eq:achievable_SE_appro_result_mrc} becomes
\begin{align}\label{eq:achievable_MRC_appro_ip_limit}
R_{{\rm{IP}},n}^{\rm{MRC}} \!\to\! {\log _2}\left( {1 \!+\!\frac{\kappa {E_u\beta_n\left(M^{\alpha}K_n\!+\!\tau E_u\beta_n\right)}}{ {M^{2\alpha\!-\!1}\left(K_n\!+\!1\right)}}}\right),\; \text{as} \; M \!\to\! \infty.
\end{align}
In contrast to the case of perfect CSI, $R_{{\rm{IP}},n}^{\rm{MRC}}$ is dependent on both the Rician $K$-factor and $\alpha$. When $K_n=0$ and $\alpha=1/2$, \eqref{eq:achievable_MRC_appro_ip_limit} tends to a constant value as
\begin{align}\label{eq:power_scaling_MRC_imperfect_1}
R_{{\rm{IP}},n}^{{\rm{MRC}}} \to {\log _2}\left( {1 +\kappa \tau {E_u^2}{\beta _n^2}}\right) ,\; \text{as} \; M \to \infty.
\end{align}
While for $\alpha=1$, \eqref{eq:achievable_MRC_appro_ip_limit} converges to a constant value as
\begin{align}\label{eq:power_scaling_MRC_imperfect_2}
R_{{\rm{IP}},n}^{{\rm{MRC}}} \!\to\! {\log _2}\left( { {1\!+\! \frac{\kappa K_n E_u \beta_n}{K_n +1 }}} \right), \;  \text{as} \; M \!\to\! \infty.
\end{align}
Note that \eqref{eq:power_scaling_MRC_imperfect_1} and \eqref{eq:power_scaling_MRC_imperfect_2} reduce to \cite[Eq. (86)]{zhang2014power} and \cite[Eq. (87)]{zhang2014power}, respectively, for infinite resolution. Both \eqref{eq:power_scaling_MRC_imperfect_1} and \eqref{eq:power_scaling_MRC_imperfect_2} clearly indicate that as $\kappa$ gets larger, a monotonic increase in the SE is achieved. Moreover, \eqref{eq:power_scaling_MRC_imperfect_2} converges to \eqref{eq:power_scaling_mrc_perfect} for very strong LoS components, i.e., $K_n \to \infty$, where the channel estimation becomes far more robust. This is because quantities in the fading matrix that were random before becomes deterministic now.
\end{rem}

\vspace{-6mm}
\section{Numerical Results}\label{se:numerical_result}
In this section, we provide numerical results to verify our derived results in Section \ref{se:uplink}. The users are assumed to be uniformly distributed in a hexagonal cell with a radius of 1000 meters, while the smallest distance between the user to the BS is $r_\text{min}= 100$ meters. Moreover, the pathloss is modeled as $r_n^{-v}$ with $r_n$ denoting the distance between the $n$th user to the BS and $v=3.8$ denoting the path loss exponent, respectively. A log-normal random variable $s_n$ with standard deviation $\delta_s = 8 $ dB is used to model the shadowing. Combining these factors, the large-scale fading is $\zeta_n = s_n (r_n/r_\text{min})^{-v}$. We further assume $\theta_n$ are uniformly distributed within the interval $[-\pi/2,\pi/2]$. Due to the space constraint, we only present the curves of perfect CSI. 

\begin{figure}[t]
\centering
\includegraphics[scale=0.5]{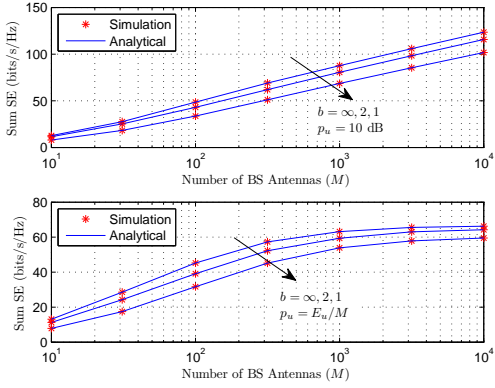}
\vspace{-2mm}
\caption{Simulated and analytical SEs of massive MIMO systems with low-resolution ADCs ($K=10$ dB and $N = 10$).
\label{fig:MRC}}
\vspace{-5mm}
\end{figure}

In Fig. \ref{fig:MRC}, the simulated uplink SE along with the proposed analytical result \eqref{eq:achievable_SE_MRC_appro_result_P} are plotted against the number of BS antennas $M$. We consider different ADC resolutions as $b=1,2, \text{and }\infty$. We find that our derived approximated results are notably matched with the simulation results for all $M$. As predicted in Section \ref{se:uplink}, we validate that a higher ADC resolution improves the SE. Moreover, with only 2 bits, the SE nearly approaches the one of ideal ADCs with infinite resolution. For fixed values of $p_u$, all curves grow without bound by increasing $M$, while for variable values of $p_u$, the corresponding curves eventually saturate with an increased $M$.

\begin{figure}[t]
\centering
\includegraphics[scale=0.485]{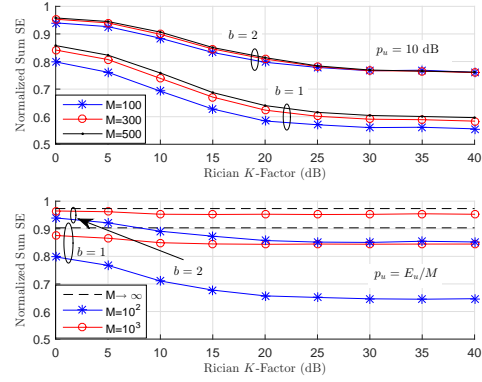}
\vspace{-2mm}
\caption{Normalized SE of massive MIMO systems with low-resolution ADCs ($N = 10$).
\label{fig:MRC_K}}
\vspace{-5mm}
\end{figure}
Figure \ref{fig:MRC_K} further illustrates the effects of the Rician $K$-factor and the number of BS antennas on the SE performance. We normalize the SE of ideal ADCs ($b \to \infty$) as the benchmark one. For fixed $p_u$, the rate loss induced by low-resolution ADCs increases as the Rician $K$-factor becomes larger. This is because quantities of fading channels that were random before become deterministic now. For variable $p_u$ and $M \to \infty$, the SE reduces to a fixed value as derived in \eqref{eq:power_scaling_mrc_perfect}. Moreover, the rate loss for fixed $p_u$ is much larger than the case of variable $p_u$, which means a larger Rician $K$-factor can increase the vulnerabilities of SE for fixed $p_u$.

\vspace{-2mm}
\section{Conclusions}\label{se:conclusion}

In this letter, we have derived both the large antenna limiting and approximated expressions for the cases of perfect and imperfect CSI, respectively. These expressions reveal how the uplink SE change with the ADC resolution, the Rician $K$-factor, and the number of BS antennas. For both perfect and imperfect CSI, the derived uplink SE limits converge to the same constant as $M \to \infty$. Finally, massive MIMO can achieve a fairly good performance with only 2-bits-resolution ADCs for a small Rician $K$-factor.

\vspace{-2mm}
\bibliographystyle{IEEEtran}
\bibliography{IEEEabrv,1_bit_Ref}

\end{document}